\documentclass[pra, twocolumn, floatfix]{revtex4}
\usepackage{graphicx}
\usepackage{color}
\usepackage{amsmath, amsfonts, amssymb, bm}
\usepackage{braket}
\begin{document}
\title{Resonant electron scattering and dielectronic recombination\\ in two-center atomic systems}
\author{A. Eckey}
\author{A. Jacob}
\author{A. B. Voitkiv}
\author{C. M\"uller}
\affiliation{Institut f\"ur Theoretische Physik I, Heinrich Heine Universit\"at D\"usseldorf, Universit\"atsstr. 1, 40225 D\"usseldorf, Germany}
\date{\today}
\begin{abstract}
Electron scattering and dielectronic recombination with an ion in the presence of a neighboring atom is studied. The incident electron is assumed to be captured by the ion, leading to resonant excitation of the atom which afterwards may stabilize either by electron re-emission or radiative decay. We show that the participation of the atom can strongly affect, both quantitatively and qualitatively, the corresponding processes of electron scattering and recombination. Various ion-atom systems and electronic transitions are considered. In particular we show that electron scattering under backward angles may be strongly enhanced and derive the scaling behavior of two-center dielectronic recombination with the principal quantum numbers of the participating atomic states.
\end{abstract}


\maketitle

\section{Introduction}

When an electron impinges on an ion, which carries at least one bound electron, a variety of different processes may occur \cite{AMueller}. In particular, for certain resonance energies, the incident electron may be captured by the ion, forming an autoionizing state. The latter can stabilize either through spontaneous radiative decay, so that the electron has eventually recombined with the ion in a bound state. Or an Auger decay, caused by electron-electron correlations, occurs, leading to re-emission of the electron which ends up in the continuum. These processes are called dielectronic recombination and resonant electron scattering, respectively.

Autoionizing transitions can also involve electrons which are located at two different atomic centers.
Penning ionization represents a famous example of such an interatomic correlation effect.
Another very interesting process driven by two-center electronic correlations is radiationless decay
of an inner-valence vacancy in one of the atoms, in case when a single-center Auger decay of this
vacancy is energetically forbidden. Then the vacancy may decay by transferring excitation energy
to a neighbor atom. This process, which is known as interatomic Coulombic decay (ICD) \cite{ICD, ICDres, ICDrev}, can still be much faster than a single-center radiative decay. It has been observed in a  variety of systems, comprising noble gas dimers \cite{dimers}, clusters \cite{clusters} and water molecules \cite{water}. The closely related process of two-center photoionization was also studied, both theoretically \cite{2CPI, Perina} and experimentally \cite{2CPIexp}.

While in most experimental studies of ICD the autoionizing state is created via photoabsorption, 
recently some experiments on interatomic processes were carried out, where the vacancy results from  electron impact. ICD following electron-impact ionization, with subsequent molecular dissociation, was observed in argon dimers and trimers for incident electron energies in the range from a few keV \cite{Lanzhou, Lanzhou2} down to about 30--100\,eV \cite{Dorn}. ICD was also identified after low-energy electron impact on water clusters adsorbed on condensed noble-gas surfaces \cite{Grieves}. 

From the side of theory, two more electron-impact induced processes were considered. On the one hand, two-center dielectronic recombination (2CDR) was studied \cite{2CDR}, where an incident electron is captured by an ion, leading to resonance excitation of a neighboring atom. The latter afterwards de-excites via spontaneous radiative decay (see Fig.~1). 2CDR thus represents a resonant electron capture process in which 
the total charge of the centers is changed. It was shown that 2CDR can dominate over the single-center process of radiative recombination by orders of magnitude. On the other hand, interatomic Coulombic electron capture (ICEC) has been calculated \cite{ICEC}. Here, the incident electron energy is so large, that the energy set free upon its capture to the ion suffices to ionize the neighboring atom. Keeping the total charge of the two centers unchanged, ICEC effectively results in an interatomic electron exchange between both centers. The process has also been investigated in a system of two quantum dots \cite{q-dots}.

\begin{figure}[b]  
\vspace{-0.25cm}
\begin{center}
\includegraphics[width=0.5\textwidth]{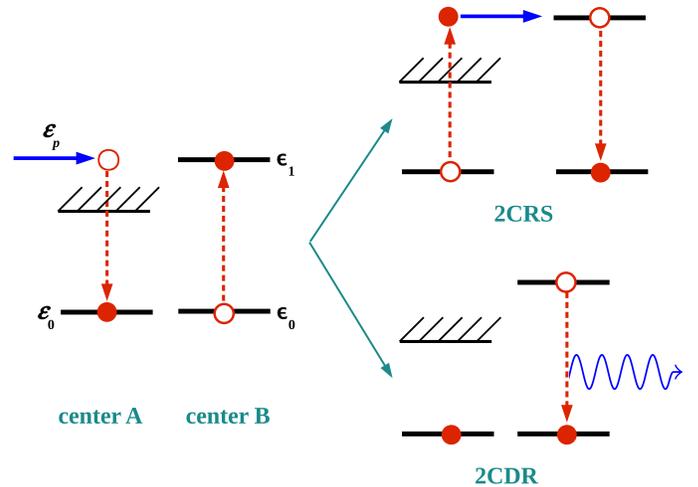}
\end{center}
\vspace{-0.5cm} 
\caption{Schemes of two-center resonant scattering (2CRS) and two-center dielectronic recombination (2CDR). An incident electron is first captured at center $A$, leading to resonant excitation at center $B$. Subsequently, this metastable state may either decay backwards (via ICD) in the case of 2CRS or radiatively stabilize by photoemission in the case of 2CDR.}
\label{figure1}
\end{figure}

In the present paper, we introduce a new example of an atomic two-center process which occurs after electron impact. The autoionizing state formed by electron capture to an ion with simultaneous excitation of a nearby atom can also decay via re-ejection of the electron. As a result, the electron has been scattered resonantly from the two-center atomic system, see Fig.~\ref{figure1}. Therefore, we call this novel process two-center resonance scattering (2CRS). It can occur for certain incident electron energies and interferes with the usual single-center scattering of the electron from the ion (Rutherford scattering). We will show that, due to 2CRS, electron-ion scattering can be qualitatively modified and strongly enhanced under backward angles. Besides, we shall also continue our investigations of 2CDR which interferes with the single-center process of radiative recombination and represents a competing process for 2CRS. The dependence of 2CDR on the quantum numbers, which characterize the auto\-ionizing state, will be derived, this way generalizing and substantially extending our earlier studies \cite{2CDR}.

Our paper is organized as follows. In Sec.~II we present our theoretical considerations of 2CRS. After formulating the general framework, closed analytical formulas for the differential cross section of this scattering process will be obtained. They include quantum interference effects and refer to the capture of the incident electron to either the 1$s$ ground state (Sec.~II.A) or the 2$s$ state (Sec.~II.B) of atom $A$.
Afterwards, in Sec.~III we turn to 2CDR, considering a generic situation where the electron is captured to an arbitrary $s$-state at center $A$, with simultaneous excitation of a general $p$-state at center $B$. The scaling of the 2CDR cross section with the principal quantum numbers of these states will be revealed. In Sec.~IV we illustrate our findings by some numerical examples and discuss their physical implications. Concluding remarks are given in Sec.~V. Atomic units (a.u.) will be used throughout unless otherwise stated.

\section{Theory of two-center resonance scattering}

In order to reveal the basic physics of 2CRS, we shall consider the process in a very elementary atomic system. We assume that the electron is incident on a proton forming the center $A$, and a hydrogen-like atom (or ion) of effective nuclear charge $Z_B$ represents the center $B$. The latter is initially in its ground state and separated from center $A$ by a distance $R$ large enough, such that one can still speak about individual atomic entities. Both the proton and the atomic nucleus are supposed to be at rest during the scattering process. We take the position of the proton as the origin of our coordinate system and denote the coordinates of the atomic nucleus, the incident electron, and the atomic electron by ${\bf R}$, ${\bf r}$ and ${\bf r}' ={\bf R} + {\boldsymbol \xi} $, respectively, where $\boldsymbol{\xi}$ is the position of the atomic electron with respect to the atomic nucleus. The $z$ axis is chosen to lie along the incident electron momentum, which also serves as our quantization axis.

In the process under consideration one has essentially three different basic two-electron configurations, which are resonant with each other and schematically illustrated in Fig.~\ref{figure1}:
(I) $\Psi_{{\bf p},0} = \varphi^{(+)}_{\bf p}({\bf r}) \chi_0(\boldsymbol{\xi})$ with total energy $E_{{\bf p}0} = \varepsilon_p + \epsilon_0$, where the incident electron is in the continuum with asymptotic momentum ${\bf p}$ and the electron of atom $B$ is in the ground state;
(II) $\Psi_{01} = \varphi_0({\bf r}) \chi_1(\boldsymbol{\xi})$ with total energy $E_{01} = \varepsilon_0 + \epsilon_1$, in which the incident electron has been captured to a bound state $\varphi_0$ of the newly formed atom $A$ while the electron of the atom $B$ is in the excited state $\chi_1$;
(III) $\Psi_{{\bf p}',0} = \varphi^{(-)}_{{\bf p}'}({\bf r}) \chi_0(\boldsymbol{\xi})$ with total energy $E_{{\bf p}'0} = \varepsilon_{p'} + \epsilon_0$, where the electron has been re-emitted into the continuum with asymptotic momentum ${\bf p}'$ and the electron of atom $B$ has returned to the ground state.

Within the second order of time-dependent perturbation theory, the probability amplitude for 2CRS can be written as
\begin{eqnarray}
S_{\chi_1} &=& -\int_{-\infty}^{\infty} dt\, \mathcal{M}(-{\bf p}')\, e^{-i(E_{01}-E_{{\bf p}'0})t}\nonumber\\
& & \times \int_{-\infty}^{t} dt'\, \mathcal{M}({\bf p})\, e^{-i(E_{{\bf p}0}-E_{01})t'}
\label{S}
\end{eqnarray}
where the subscript indicates the excited state in atom $B$ which is involved. The matrix elements are given by 
\begin{eqnarray} 
\mathcal{M}({\bf p}) = \langle \Psi_{01}|\hat{V}_{AB}| \Psi_{{\bf p},0} \rangle
\end{eqnarray}
and an according expression for $\mathcal{M}(-{\bf p}')$. Note in this regard that 
we shall use continuum states which are normalized to a quantization volume of unity and fulfill the relation $\big[\varphi^{(-)}_{{\bf p}'}({\bf r})\big]^*=\varphi^{(+)}_{-{\bf p}'}({\bf r})$ \cite{LandauQM}.
The two-center interaction between the electrons is given by
\begin{eqnarray} 
{\hat V}_{AB} = \frac{ {\bf r}\cdot {\bm \xi} }{ R^3 }
 - \frac{ 3({\bf r}\cdot{\bf R})({\bm \xi}\cdot{\bf R})}{ R^5 }
\label{VAB}
\end{eqnarray}
where a dipole-allowed transition in atom $B$ is assumed and retardation effects are neglected.
After performing the time integration in Eq.~\eqref{S}, we arrive at
\begin{eqnarray}
S_{\chi_1} = -2\pi i\, \delta(\varepsilon_{p'}-\varepsilon_p)\,\frac{\mathcal{M}(-{\bf p}')\mathcal{M}({\bf p})}{\varepsilon_p+\epsilon_0-\varepsilon_0-\epsilon_1+\frac{i}{2}\Gamma}
\label{S2}
\end{eqnarray}
where $\Gamma=\Gamma_r+\Gamma_a$ denotes the total width of the excited state $\chi_1$ in atom $B$. It has been inserted to account for the finite lifetime of this state and consists of the radiative width $\Gamma_r$ and the two-center Auger width $\Gamma_a$. The latter is given by
\begin{eqnarray}
\Gamma_a = \frac{p}{(2\pi )^2}\int d\Omega\, \big| \mathcal{M}(-{\bf p}') \big|^2
\label{Gamma_Auger}
\end{eqnarray}
where $\mathcal{M}(-{\bf p}')$ is evaluated with $p'=p$. The delta function in Eq.~\eqref{S2} displays the law of energy conservation in the process, which enforces an elastic scattering of the electron under the assumptions made.

From Eq.~\eqref{S2} one can obtain the differential scattering cross section in the usual way, by taking the absolute square of the amplitude, dividing it by the interaction time and the incident electron flux $j=p$, and integrating the resulting expression over the final electron energy. This leads to
\begin{eqnarray}
\frac{d\sigma_{\chi_1}}{d\Omega} = 
\frac{|\mathcal{M}(-{\bf p}')|^2\,|\mathcal{M}({\bf p})|^2}{(2\pi)^2 \left(\Delta^2 + \frac{1}{4}\Gamma^2\right)}
\label{CS}
\end{eqnarray}
where the absolute value of ${\bf p}'$ is fixed by energy conservation to $p'=p$ and the detuning from the resonance $\Delta = \varepsilon_p+\epsilon_0-\varepsilon_0-\epsilon_1$ has been introduced. By performing the integral over the scattering angles using Eq.~\eqref{Gamma_Auger}, one obtains the total cross section
\begin{eqnarray}
\sigma_{\chi_1} = 
\frac{\Gamma_a\,|\mathcal{M}({\bf p})|^2}{p\left(\Delta^2 + \frac{1}{4}\Gamma^2\right)}\ .
\label{CS_tot}
\end{eqnarray}

\subsection{Capture to the 1s state}
\label{1s}

In order to evaluate the differential cross section \eqref{CS} we need to specify the electron states. The incoming and outgoing electron states in the continuum will always be described by the corresponding ``in'' and ``out'' Coulomb waves with asymptotic momenta ${\bf p}$ and ${\bf p}'$, respectively (see, e.g., \cite{LandauQM}). Moreover, in this section, we will assume that the incident electron is captured to the $1s$ ground state of the hydrogen atom formed at center $A$. The one-electron atom at center $B$ is simultaneously excited from the $1s$ to a $2p_m$ state. The corresponding matrix elements can be calculated by standard means. 

For example, if the excited state in atom $B$ is the $2p_0$ state, one finds
\begin{eqnarray}
\mathcal{M}({\bf p}) &=& \frac{2^{9}\pi i\, a_B\,Z_A^3\,e^{-2\nu \text{arccot}(\nu)}}{3^5\,(\varepsilon_p-\varepsilon_0)^3}\sqrt{\frac{p\,(1+\nu^2)}{1-e^{-2\pi \nu}}} \notag \\
&  & \times\left(\frac{n_z}{R^3}-\frac{3({\bf n}\cdot {\bf R})R_z}{R^5}\right) e^{i\delta}\ ,
\label{M}
\end{eqnarray}
with the phase $\delta = -\arctan(\nu) + \mbox{arg}[\Gamma(1-i\nu)]$.
Here, $Z_A$ is the nuclear charge of the ion at center $A$ ($Z_A=1$ for a proton), $\nu = Z_A/p$ the Sommerfeld parameter, $a_B=1/Z_B$ the Bohr radius in atom $B$, and ${\bf n} = {\bf p}/p$ is a unit vector along the electron momentum with components $n_x$, $n_y$ and $n_z$.
An analogous expression holds for the matrix element $\mathcal{M}(-{\bf p}')$ which contains the outgoing momentum ${\bf p}'$. The corresponding unit vector ${\bf n'}={\bf p'}/p$ can be parametrized by the scattering angles according to $n_x'=\sin\vartheta\cos\phi$, $n_y'=\sin\vartheta\sin\phi$ and $n_z'=\cos\vartheta$.

Let us assume for a moment that the interatomic separation vector is parallel to the incident electron momentum, ${\bf R} = R{\bf e}_z$. Then only the $2p_0$ state in atom $B$ contributes to 2CRS. By inserting Eq.~\eqref{M} into Eq.~\eqref{CS}, we obtain the corresponding cross section
\begin{eqnarray}
\frac{d\sigma_{2p_0}}{d\Omega} &=& 
\frac{2^{40}\pi^2\,a_B^4\, Z_A^{10}\,\nu^2\,e^{-8\nu \text{arccot}(\nu)}}{3^{20} R^{12} (1-e^{-2\pi \nu})^2\,(\varepsilon_p-\varepsilon_0)^{10} }\nonumber\\
& & \times\, \frac{1}{\Delta^2 + \frac{1}{4}\Gamma^2}\,\cos^2\!\vartheta\ .
\label{CSz}
\end{eqnarray}
From this expression, we can read off some interesting properties of 2CRS. (i) It shows a very different dependence on the scattering angle than the competing single-center process of Rutherford scattering, whose famous cross section reads
\begin{eqnarray}
\frac{d\sigma_{\rm R}}{d\Omega} = \left(\frac{Z_A}{4 \varepsilon_p}\right)^{\! 2} \frac{1}{\sin^4\!\left(\frac{\vartheta}{2}\right)}\ .
\label{CS_R}
\end{eqnarray}
The angular dependence in Eq.~\eqref{CSz} has a characteristic dipole form, similarly to photoionization from the ground state by a photon which is polarized along the $z$ direction. (ii) While the two-center processes of photoionization \cite{2CPI} and dielectronic recombination \cite{2CDR} scale with the internuclear distance like $R^{-6}$, 2CRS exhibits a much steeper dependence with $R^{-12}$ (in the range where $\Gamma_r\gg\Gamma_a$). This is due to the fact that 2CRS is of second order in the interatomic interaction \eqref{VAB}. 
(iii) There is also a very strong dependence on the transition energy $\varepsilon_p-\varepsilon_0$. Note in this context that also the level width $\Gamma$ depends on the transition energy in atom $B$; one typically has, for example, $\Gamma_r \sim (\epsilon_1-\epsilon_0)^3$, with $\epsilon_1-\epsilon_0\approx\varepsilon_p-\varepsilon_0$ close to the resonance. Therefore, low transition energies are generally favorable for 2CRS.
(iv) For Rutherford scattering it is well known that a calculation within the first Born approximation using plane waves for the continuum electron states gives the same cross section as the exact calculation based on Coulomb waves. For 2CRS, however, the result from a plane-wave approach differs substantially from Eq.~\eqref{CSz} \cite{Eckey}.

Assuming that the resonance condition is exactly met, $\varepsilon_p-\varepsilon_0 = \epsilon_1 - \epsilon_0$, and that $\Gamma_r = \frac{2^{17}a_B^2}{3^{11}c^3}(\epsilon_1-\epsilon_0)^3\gg\Gamma_a$ holds (which is valid for sufficiently large values of $R$), we find for the ratio of 2CRS to Rutherford scattering the expression
\begin{eqnarray}
\frac{d\sigma_{2p_0}/d\Omega}{d\sigma_{\rm R}/d\Omega} &=& 
\frac{2^{12}\,3^{2}\,\pi^2\,c^6\,Z_A^8\,\varepsilon_p^2\,\nu^2\,e^{-8\nu \text{arccot}(\nu)}}{R^{12} (1-e^{-2\pi \nu})^2\,(\varepsilon_p-\varepsilon_0)^{16} }\nonumber\\
& & \times \cos^2\!\vartheta\,\sin^4\!\left(\frac{\vartheta}{2}\right)
\label{ratio1s_Gamma_r}
\end{eqnarray}
which is maximized under backward angles ($\vartheta=\pi$). 
For instance, assuming an incident energy of $\varepsilon_p=1$\,a.u., a binding energy of $\varepsilon_0=-0.5$\,a.u. and an interatomic distance of $R=10$\,a.u., formula \eqref{ratio1s_Gamma_r} predicts that the ratio reaches a maximum value of about 20. Note that the chosen energies fit to resonant electron scattering from a proton in the presence of a neighboring He$^+$ ion. At $R=10$\,a.u., one has $\Gamma_r\approx 2 \Gamma_a$ in this system which is not quite sufficient to apply Eq.~\eqref{ratio1s_Gamma_r}, though. Rather, the ratio is reduced to about 10.

In the opposite limit $\Gamma_r\ll \Gamma_a$, we obtain from Eqs.~\eqref{CS} and \eqref{Gamma_Auger} that the 2CRS cross section on the resonance becomes independent of $R$, this way forming a plateau region where the cross section attains the value $d\sigma_{2p_0}/d\Omega = (3/p)^2\,\cos^2\vartheta$. Its ratio with the Rutherford cross section for backward scattering then simply reads $(d\sigma_{2p_0}/d\Omega)/(d\sigma_{\rm R}/d\Omega) = 72\,\varepsilon_p / Z_A^2$.

The generalization of Eq.~\eqref{CSz} to interatomic distance vectors $\bf R$ of arbitrary direction is straightforward \cite{Eckey}. Note that in the general case also the excited $2p_{+1}$ and $2p_{-1}$ states in atom $B$ need to be taken into account. The corresponding transition amplitudes $S_{2p_0}$, $S_{2p_1}$ and $S_{2p_{-1}}$ have to be added coherently. The result is
\begin{eqnarray}
\frac{d\sigma_{2p}}{d\Omega}&= & 
\frac{2^{36}\pi^2\,a_B^4\,Z_A^{10}\,\nu^2\,e^{-8\nu \text{arccot}(\nu)}}{3^{20} R^{12} (1-e^{-2\pi \nu})^2\,(\varepsilon_p-\varepsilon_0)^{10}}\nonumber\\
& & \times\, \frac{1}{\Delta^2 + \frac{1}{4}\Gamma^2}\,
\Big( B_0(\vartheta,\phi) + \mathfrak{R}[B_1(\vartheta,\phi)] \Big)^2\ 
\label{sigma_2CRS_2p}
\end{eqnarray}
with the angle- and geometry-dependent functions
\begin{eqnarray}
B_0(\vartheta,\phi) &=& \left(1-3\rho_z^2\right)\left[ n_z' - 3({\bf n}'\cdot{\boldsymbol\rho})\rho_z\right]\ ,\nonumber\\ 
B_1(\vartheta,\phi) &=& 3\rho_z\rho_-\left[n_+' - 3({\bf n}'\cdot{\boldsymbol\rho})\rho_+\right]\ ,
\label{B01}
\end{eqnarray}
where $n_+' = n_x'+ i n_y'=\sin\vartheta\, e^{i\phi}$ and $\boldsymbol{\rho}={\bf R}/R$ is a unit vector with components $\rho_x$, $\rho_y$, $\rho_z$, and $\rho_\pm = \rho_x \pm i\rho_y$. Let us give some examples. When $\boldsymbol{\rho}\perp{\bf e}_z$, we obtain $B_0(\vartheta,\phi)=\cos\vartheta$, $B_1(\vartheta,\phi)=0$. The cross section then exhibits a dependence on $\cos^2\!\vartheta$ like in Eq.~\eqref{CSz}, but the total value is reduced by a relative factor of 4. When instead $\boldsymbol{\rho}=(\sqrt{2},0,1)/\sqrt{3}$, then $B_0(\vartheta,\phi)=0$ and $\mathfrak{R}[B_1(\vartheta,\phi)]=-(2\cos\vartheta+\sqrt{2}\,\sin\vartheta\cos\phi)$. In this case, the cross section depends also on the azimuthal angle.

In a complete picture of the scattering process, the 2CRS channel must be considered jointly with the single-center electron-ion scattering, since both lead to the same final state. Thus, the three amplitudes \eqref{S} for excited $2p_0$ and $2p_{\pm 1}$ states must be added coherently to the amplitude for Rutherford scattering
\begin{eqnarray}
S_{\rm R} = \frac{i \pi^2 Z_A}{\varepsilon_p \sin^2(\frac{\vartheta}{2})}\,e^{i\delta_{\rm R}}\,\delta(\varepsilon_p-\varepsilon_{p'})
\end{eqnarray}
with the phase 
\begin{eqnarray}
\delta_{\rm R} = \nu\ln\left[\sin^2\!\left(\frac{\vartheta}{2}\right)\right]+2\arg\left[ \Gamma\left(1-i\nu\right) \right]\ .
\end{eqnarray}
This way, quantum interference between the single-center and two-center pathways of scattering is accounted for. The differential cross section thus becomes 
\begin{eqnarray}
\frac{d\sigma}{d\Omega} = \frac{1}{T}\int_{0}^{\infty}\frac{dp'\,p'}{(2\pi)^3}\,
\big|S_{\rm R}+S_{2p_0}+S_{2p_1}+S_{2p_{-1}}\big|^2
\end{eqnarray}
where $T$ denotes the interaction time. By performing the necessary calculational steps, we obtain
\begin{eqnarray}
\frac{d\sigma}{d\Omega} &=& \frac{1}{\Delta^2 + \frac{1}{4}\Gamma^2}\,
\Bigl\{ \big[\Delta - C(\vartheta,\phi)\cos\Phi(\vartheta)\big]^2 \nonumber\\
& & +\big[\frac{1}{2} \Gamma+C(\vartheta,\phi)\sin\Phi(\vartheta)\big]^2 \Bigr\}
\,\frac{d\sigma_{\rm R}}{d\Omega}\ .
\label{CS_int}
\end{eqnarray}
For the sake of a compact notation, we have introduced here the interference phase 
\begin{eqnarray}
\Phi(\vartheta) = \delta_{\rm R}-2\delta = \nu\ln\left[\sin^2\!\left(\frac{\vartheta}{2}\right)\right]+2\arctan(\nu)
\end{eqnarray}
and the angle- and geometry-dependent function
\begin{eqnarray}
C(\vartheta,\phi) = A(\vartheta)\big\{ B_0(\vartheta,\phi) + \mathfrak{R}[B_1(\vartheta,\phi)]\, \big\}\ ,
\label{C}
\end{eqnarray}
with $B_{0,1}(\vartheta,\phi)$ given in Eq.~\eqref{B01} and
\begin{eqnarray}
A(\vartheta) =  \frac{2^{20}\pi\, a_B^2\,Z_A^4\, \varepsilon_p\, \nu\,e^{-4\nu \text{arccot}(\nu)}}{3^{10} R^6 (1-e^{-2\pi\nu})(\varepsilon_p-\varepsilon_0)^5}\, \sin^2\!\left(\frac{\vartheta}{2}\right)\ .
\end{eqnarray}

\subsection{Capture to the 2s state}
In Sec.~\ref{1s} we have seen that the cross section of 2CRS shows a very strong dependence on the electronic transition energy. This renders the consideration of a hydrogen atom ($Z_B=1$) -- rather than a hydrogen-like ion -- located at center $B$ interesting. For energetic reasons, though, the process discussed in Sec.~\ref{1s} cannot take place if hydrogen constitutes the neighboring atom. Because the energy set free by capturing the incident electron at center $A$ would exceed any transition energy to a bound state at center $B$ and, thus, would lead to ionization of atom $B$ via ICEC rather than to 2CRS. However, if the proton captures the incident electron not into the ground state, but instead into an excited state of the hydrogen atom at center $A$, then 2CRS can proceed in the presence of a neighboring hydrogen atom at center $B$. 

For definiteness, we will assume in this section, that the electron is captured to the $2s$ state. If the capture is accompanied by resonant excitation from $1s$ to $2p_0$ at center $B$, the corresponding matrix element reads
\begin{eqnarray}
\mathcal{M}({\bf p}) &=& \frac{2^7 \sqrt{2}\, \pi i}{3^5}\frac{e^{-2\nu \text{arccot}(\nu/2)}}{(\varepsilon_p-\varepsilon_0)^3} 
\sqrt{\frac{1+\nu^2}{\nu\left(1-e^{-2\pi\nu}\right)}} 
\notag \\
&  & \times\left(\frac{n_z}{R^3}-\frac{3({\bf n}\cdot {\bf R})R_z}{R^5}\right) e^{i\delta}
\end{eqnarray}
where $\varepsilon_0$ now denotes the energy of the $2s$ state in atom $A$.
Restricting ourselves to the case where ${\bf R} = R{\bf e}_z$, we find the cross section to be
\begin{eqnarray}
\frac{d\sigma_{2p_0}}{d\Omega}&= &\frac{2^{32} \pi^2 }{3^{20} R^{12} }\frac{(1+\nu^2)^2\, e^{-8\nu \text{arccot}(\nu/2)}}{\nu^2\left(1-e^{-2\pi\nu}\right)^2(\varepsilon_p-\varepsilon_0)^{12}} \notag \\
&  &\times\,\frac{1}{\Delta^2 + \frac{1}{4}\Gamma^2} \, \cos^2\!\vartheta\ .
\end{eqnarray}

If the resonance condition is exactly fulfilled, $\varepsilon_p-\varepsilon_0 = \epsilon_1 - \epsilon_0$, and $\Gamma_r \gg \Gamma_a$ holds (which is valid for $R\gtrsim 30$\,a.u.), we find for the ratio of 2CRS to Rutherford scattering under backward angles the expression
\begin{eqnarray}
\frac{d\sigma_{2p_0}/d\Omega}{d\sigma_{\rm R}/d\Omega}\biggl|_{\vartheta=\pi} =
\frac{2^{2}\,3^{2}\,\pi^2\,c^6\,\varepsilon_p^2\,(1+\nu^2)^2\,e^{-8\nu \text{arccot}(\nu/2)}}{R^{12}\,\nu^2\, (1-e^{-2\pi \nu})^2\,(\varepsilon_p-\varepsilon_0)^{18} }
\label{ratio2s}
\end{eqnarray}
Assuming an incident energy of $\varepsilon_p=0.25$\,a.u., a binding energy of $\varepsilon_0=-0.125$\,a.u. and an interatomic distance of $R=30$\,a.u., this ratio reaches a value of about one. We note that, in comparison with the example given below Eq.~\eqref{ratio1s_Gamma_r}, the amplyfing effect from the smaller transition energy in the present case is counteracted by the larger interatomic distance required to guarantee $\Gamma_r \gg \Gamma_a$. In the opposite limit $\Gamma_r \ll \Gamma_a$, the corresponding formula given in Sec.~\ref{1s} holds.

Taking in a complete picture the interference of the amplitudes $S_{2p_0}$ and $S_{\rm R}$ into account, we arrive at the cross section
\begin{eqnarray}
\frac{d\sigma}{d\Omega} &=& \frac{1}{\Delta^2 + \frac{1}{4}\Gamma^2}\,
\Bigl\{ \big[\Delta - D(\vartheta)\cos\Phi(\vartheta)\big]^2 \nonumber\\
& & +\big[\frac{1}{2} \Gamma+D(\vartheta)\sin\Phi(\vartheta)\big]^2 \Bigr\}
\,\frac{d\sigma_{\rm R}}{d\Omega}\ .
\end{eqnarray}
where the angle-dependent function 
\begin{eqnarray}
D(\vartheta) = \frac{2^{18}\pi\, \varepsilon_p (1+\nu^2)\,e^{-4\nu \text{arccot}(\nu/2)} \sin^2(\frac{\vartheta}{2})\cos\vartheta}{3^{10} R^6\, \nu(1-e^{-2\pi \nu}) (\varepsilon_p-\varepsilon_1)^6} 
\end{eqnarray}
has been introduced.

Before moving on to the next section, we point out that in an even more complete picture of 2CRS one should also take into account the elastic scattering of the incident electron on atom $B$. The corresponding probability amplitude needs to be included in the coherent summation. However, one can estimate that in the case of a neutral atom at center $B$ (rather than an ion), the influence of the additional scattering channel will not change our results dramatically. Let us consider as a specific example a hydrogen atom at center $B$. The cross section for scattering from this center then is $d\sigma_B/d\Omega = 4(q^2+8)^2/(q^2+4)^4$, where $q=2p\sin(\frac{\vartheta}{2})$ denotes the momentum transfer. 2CRS is relatively strongest under backward angles. In the situation of the present Sec.~II.B the resonant incident momentum is $p=1/\sqrt{2}$\,a.u., corresponding to $q^2=2$\,a.u. for $\vartheta=\pi$. Thus, $d\sigma_B/d\Omega \approx 0.3\,\mbox{a.u.}\approx d\sigma_{\rm R}/d\Omega$. For parameters, where the 2CRS cross section largely exceeds the Rutherford cross section under backward angles, we therefore expect that the omission of elastic scattering from atom $B$ does not change our predictions for this angular region substantially. We plan to investigate this aspect in more detail in a forthcoming study.

\section{Theory of two-center dielectronic recombination}

The autoionizing state $\Psi_{01}$, that has been formed as an intermediate state during 2CRS, can also stabilize through photoemission. Then the two-center system ends up in the state $\Psi_{00}=\varphi_0({\bf r}) \chi_0(\boldsymbol{\xi})$ with total energy $E_{00} = \varepsilon_0 + \epsilon_0$ and both electrons in their corresponding ground states. In this case, the incident electron has recombined with the atom $A$, and the process represents a two-center version of dielectronic recombination (2CDR).

In addition to the electrons, the quantum degrees of freedom of 2CDR -- being a radiative process -- are also represented by the radiation field. It is initially in its vacuum state $|0\rangle$ and during the process undergoes a transition into the final state $|{\bf k},\lambda\rangle$, describing the emission of a photon with momentum $\bf k$ and polarization vector ${\bf e}_{{\bf k}\lambda}$. 

Thus, within the second order of time-dependent perturbation theory, the probability amplitude for 2CDR can be formulated as [cp. Eq.~\eqref{S}]
\begin{eqnarray}
\mathcal{S}{\chi_1} &=& - \int_{-\infty}^\infty dt\, \langle \Psi_{00} | \hat{V}_\gamma | \Psi_{01} \rangle\,  e^{- i ( E_{01} - E_{00} ) t } \nonumber\\
& & \times\,\int_{-\infty}^t dt' \, \mathcal{M}({\bf p})\,  e^{- i ( E_{{\bf p}0} - E_{01} ) t' } .
\label{S_2CDR}
\end{eqnarray}
Here, the photonic degrees of freedom have already been evaluated and the interaction responsible for the photo\-emission is expressed in the effective form
\begin{eqnarray}
\hat{V}_{\gamma} = \frac{1}{c} {\bf A}_{{\bf k}\lambda} \cdot \hat{\bf p}_B ~,
\label{V_gamma}
\end{eqnarray}
where ${\bf A}_{{\bf k}\lambda}=\sqrt{2\pi c^2/\omega_k}\,{\bf e}_{{\bf k}\lambda}\,e^{i\omega_kt}$, with $\omega_k=|{\bf k}|c$, is taken in the dipole approximation and $\hat{\bf p}_B$ denotes the momentum operator of the electron at center $B$.

In our previous studies \cite{2CDR} we considered 2CDR with capture of the electron to the $1s$ ground state in atom $A$ and excitation from $1s\to 2p$ in atom $B$. Here, we generalize our results to the case where the incident electron is captured to a general $n_As$ state in atom $A$ with simultaneous excitation of the electron in atom $B$ from the ground state to an $n_Bp$ state. Our goal is to reveal the dependencies of 2CDR on the principal quantum numbers $n_A$ and $n_B$. 

As before, we will assume that the asymptotic momentum of the incident electron lies along the $z$ axis, ${\bf n}={\bf e}_z$. If the electron capture to the $n_As$ state is accompanied by resonant excitation to the $n_Bp_0$ level, the corresponding matrix element is
\begin{eqnarray}
\mathcal{M}({\bf p}) &=& \frac{2^{10}\sqrt{2}\,\pi\,i\,Z_A^3}{\sqrt{3}\,Z_B\,R^3} \sqrt{\frac{n_B^5 (n_B-1)}{n_A^5(n_B+1)^9} \, \frac{(1+\nu^2)}{(1-e^{-2\pi\nu})} } \nonumber\\
& &  \times
\frac{p}{\varepsilon_0 - \varepsilon_p} \, \mathcal{A}\,F(2-n_B,5,4,\frac{2}{1+n_B})\,e^{i\delta}\nonumber\\
& &  \times\,(1-3\rho_z^2)
~. \label{M_gen}
\end{eqnarray}
Here, $F$ denotes the hypergeometric function \cite{LandauQM} and we have introduced the complex quantity
\begin{eqnarray}
\mathcal{A} &=& \frac{\beta_+^{\gamma+\gamma'-4}}{(-\beta_-)^{\gamma} \beta_-^{\gamma'}}
\bigg\{ \frac{1}{\beta^2} \Big[ \beta_+\beta_- F(\gamma, \gamma'-1, 2, \zeta) \nonumber\\
&\quad& + 2\beta_-^2 F(\gamma-1, \gamma'-1, 2, \zeta) \nonumber\\
&\quad& + \beta_+^{-1} \beta_-^3 F(\gamma-2, \gamma'-1, 2, \zeta) \Big] + 2\frac{n_A-1}{\beta} \nonumber\\
&\quad&  \times\Big[ \beta_+ F(\gamma, \gamma', 3, \zeta) + \beta_- F(\gamma-1, \gamma', 3, \zeta) \Big] \bigg\}\ 
\end{eqnarray}
with $\beta=2ip$, $\beta'=2Z_A/n_A$, $\beta_\pm=\beta\pm\beta'$, $\gamma=2+i\nu$, $\gamma'=2-n_A$ and the argument $\zeta = 4\beta \beta'/\beta_-^2$. Equation~\eqref{M_gen} represents a generalization of Eq.~\eqref{M} to arbitrary values of $n_A$ and $n_B$. The other matrix element in Eq.~\eqref{S_2CDR} has the form of an ordinary single-center dipole matrix element, 
\begin{eqnarray}
\langle \Psi_{00} | \hat{V}_\gamma | \Psi_{01} \rangle &=& \sqrt{\frac{2\pi}{\omega_k}}\,e^{i\omega_kt}
\langle \chi_0 |\, {\bf e}_{{\bf k}\lambda}\cdot \hat{\bf p}_B | \chi_1 \rangle \nonumber\\
&=& i \frac{2^5 \sqrt{2\pi}}{\sqrt{3\,\omega_k}\,Z_B} \sqrt{\frac{n_B^5 (n_B-1)}{ (n_B+1)^9}} (\epsilon_1 - \epsilon_0 ) e^{i\omega_k t} \nonumber\\
& &  \times  F(2-n_B,5,4,\frac{2}{1+n_B})({\bf e}_{{\bf k}\lambda}\cdot {\bf e}_z)\ \label{M_B}
\end{eqnarray}
After performing the time integrals in Eq.~\eqref{S_2CDR}, the probability amplitude can be obtained from Eqs.~\eqref{M_gen}-\eqref{M_B} according to
\begin{eqnarray}
\mathcal{S}_{n_Bp_0} = 2\pi\, \delta(\varepsilon_p-\varepsilon_0-\omega_k)\,\frac{\big|\langle \chi_0 | \hat{V}_\gamma | \chi_1 \rangle\big| \, \mathcal{M}({\bf p})}{\Delta+\frac{i}{2}\Gamma}
\label{S_2CDR_2}
\end{eqnarray}
from which the cross section follows as 
\begin{eqnarray}
\sigma_{n_Bp_0} &=& \frac{1}{jT}\int \frac{d^3k}{(2\pi)^3}\sum_{\lambda=1,2} \big| \mathcal{S}_{n_Bp_0} \big|^2 \nonumber\\
&=& \frac{\Gamma_r^{(1s)}\,|\mathcal{M}({\bf p})|^2}{p\left(\Delta^2+\frac{1}{4}\Gamma^2\right)}\ .
\label{sigma_2CDR_0}
\end{eqnarray}
It gives the only contribution to 2CDR in the special case when the atoms are spatially separated along the $z$ axis (i.e., parallel to the incident electron momentum). The integration over the emitted photon momentum and the sum over its polarizations has been expressed with the help of the partial radiative width $\Gamma_r^{(1s)}$ for decay of the excited level to the ground state. It can be written explicitly as
\begin{eqnarray}
\Gamma_r^{(1s)} &=& \frac{2^{12} (\epsilon_1-\epsilon_0)^3}{3^2 Z_B^2 c^3} \frac{n_B^5 (n_B-1)}{(n_B+1)^{9}} \nonumber\\
& & \times\, F^2(2-n_B,5,4,\frac{2}{1+n_B})
\label{Gamma_rad}
\end{eqnarray}
and shows, to leading order, a scaling behavior with the principal quantum number $n_B$ like $\Gamma_r^{(1s)}\sim n_B^{-3}$. The same holds for $|\mathcal{M}({\bf p})|^2\sim n_B^{-3}$.
The generic structure of Eq.~\eqref{sigma_2CDR_0} will prove useful for interpreting our numerical results in the next section.

By plugging Eqs.~\eqref{M_gen} and \eqref{Gamma_rad} into Eq.~\eqref{sigma_2CDR_0}, we obtain
\begin{eqnarray}
\sigma_{n_Bp_0} &=& 
\frac{2^{33} \pi^2 Z_A^6}{3^3 c^3 Z_B^4 R^6} \frac{n_B^{10} (n_B-1)^2}{n_A^5 (n_B+1)^{18}}\,\frac{1+\nu^2}{1-e^{-2\pi\nu}}\,|\mathcal{A}|^2 \nonumber\\
& &  \times\, \frac{( \epsilon_1 - \epsilon_0 )^2}{\varepsilon_p - \varepsilon_0} \frac{F^4(2-n_B,5,4,\frac{2}{1+n_B})}{\Delta^2 + \frac{1}{4}\Gamma^2}\,\mathcal{B}_0({\bf R})\ , \nonumber\\
\label{sigma_2CDR_0_final}
\end{eqnarray}
with the geometry-dependence $\mathcal{B}_0({\bf R})=(1-3\rho_z^2)^2$, which reduces to $\mathcal{B}_0({\bf R})=4$ for ${\bf R}=R{\bf e}_z$. This formula allows us to reveal the potential significance of 2CDR. The process co-occurs with single-center radiative recombination (without participation of center $B$), where the electron is captured at center $A$ and the excess energy is released via photoemission. The corresponding transition amplitude
\begin{eqnarray}
\mathcal{S}_{\rm RR} = -\frac{i}{c}\int_{-\infty}^\infty dt\, \langle \varphi_0 | {\bf A}_{{\bf k}\lambda} \cdot \hat{\bf p}_A | \varphi_{\bf p}^{(+)} \rangle\,  e^{- i ( \varepsilon_p - \varepsilon_0 ) t }
\end{eqnarray}
leads to the following cross section for radiative recombination into the $n_As$ state
\begin{eqnarray}
\sigma_{\rm RR} = \frac{2^{13}\pi^2 Z_A^6(1+\nu^2)(\varepsilon_p-\varepsilon_0)}{3c^3 n_A^5 (1-e^{-2\pi\nu})} |\mathcal{A}|^2 ~. \label{CS_RR}
\end{eqnarray}
Assuming that the resonance condition is met exactly, the ratio of the two-center and single-center recombination cross sections reads 
\begin{eqnarray}
\frac{\sigma_{n_Bp_0}}{\sigma_{\rm RR}} =
\frac{2^{24}n_B^{10} (n_B-1)^2 F^4(2-n_B,5,4,\frac{2}{1+n_B})}{3^2 Z_B^4 R^6 (n_B+1)^{18}\,\Gamma^2 }\ .
\label{2CDR_ratio}
\end{eqnarray}
This ratio can be very large, as will be discussed in Sec.~IV. In the special case $n_B=2$, and assuming that $\Gamma_r\gg \Gamma_a$ holds, it reduces to the compact formula
\begin{eqnarray}
\frac{\sigma_{2p_0}}{\sigma_{\rm RR}} = \frac{3^2 c^6}{R^6 (\epsilon_1-\epsilon_0)^6}\ .
\label{2CDR_ratio_2p}
\end{eqnarray}
which is known from our previous studies \cite{2CDR}. As an example, suppose that the transition energy is $\epsilon_1-\epsilon_0=1.5$\,a.u. and $R=10$\,a.u., then the ratio is of order 10$^7$.

Equations~\eqref{2CDR_ratio} and \eqref{2CDR_ratio_2p} compare the cross sections for the separate pathways of 2CDR versus radiative recombination. We point out that, in a complete picture of the recombination process, one needs to sum the corresponding probability amplitudes coherently, this way accounting for interference between both processes \cite{2CDR}. However, as we have just seen, close to the resonance and for interatomic distances up to few nanometers, 2CDR can dominate over the single-center process of radiative recombination by several orders of magnitude. Therefore, we shall focus in this paper on the direct channel of 2CDR and omit the interference effects.

The total level width $\Gamma=\Gamma_r + \Gamma_a^{(0)}$ in Eq.~\eqref{sigma_2CDR_0_final} consists of the radiative width and the Auger width [see Eq.~\eqref{Gamma_Auger}] of the autoionizing state which involves the $n_Bp_0$ state. The radiative width itself is composed of several contributions each of which corresponding to the spontaneous radiative decay of the excited $n_Bp$ state to a lower lying level. The main contributions come from decays to lower lying $s$ states, so that we may write approximately
\begin{eqnarray}
\Gamma_r \approx \Gamma_r^{(1s)} + \Gamma_r^{(2s)} + \Gamma_r^{(3s)} + \ldots
\end{eqnarray}
with the partial width for decay to the ground state $\Gamma_r^{(1s)}$ being largest, followed by $\Gamma_r^{(2s)}$ and so on. By inspection of the atomic database \cite{NIST} we found the approximate relations $\Gamma_r^{(2s)}\approx \frac{1}{7}\Gamma_r^{(1s)}$ and $\Gamma_r^{(3s)}\approx \frac{1}{3}\Gamma_r^{(2s)}$, which were validated for He$^+$ ions and $n_B\le 6$. Accordingly, in our numerical calculations we will approximate the radiative width for simplicity by the formula
\begin{eqnarray}
\Gamma_r \approx \Gamma_r^{(1s)} \left( 1 + \frac{1}{7} + \frac{1}{21}\right)
\label{Gamma_rad_total}
\end{eqnarray}
for excited states with $n_B\ge 4$. For lower excitations, the sum in Eq.~\eqref{Gamma_rad_total} will be truncated accordingly. 
The two-center Auger width in Eq.~\eqref{sigma_2CDR_0_final} reads
\begin{eqnarray}
\Gamma_a^{(0)} &=& \frac{2^{21} \pi Z_A^6}{3^2 Z_B^2 R^6} \frac{n_B^5 (n_B-1)}{n_A^5(n_B+1)^{9}}\, \frac{1+\nu^2}{1-e^{-2\pi\nu}}\, \frac{p^2}{ (\varepsilon_0 - \varepsilon_p)^2} \nonumber\\
& &  \times F^2(2-n_B,5,4,\frac{2}{1+n_B})\, |\mathcal{A}|^2\,(1-3\rho_z^2)^2\
\label{Gamma_Auger_0}
\end{eqnarray}
where the upper index indicates the magnetic quantum number of the excited $n_Bp_m$ state.
To leading order, the two-center Auger width also scales like $\Gamma_a^{(0)} \sim n_B^{-3}$.

So far, we have assumed that the atoms are spatially separated along the $z$ axis, ${\bf R}=R{\bf e}_z$.
For an arbitrary orientation of the interatomic separation vector ${\bf R}$, in general all $n_Bp_m$ ($m\in\{0,\pm1\}$) states contribute to 2CDR. The total cross section -- ignoring interference with radiative recombination -- can be written as the sum over the respective contributions, according to 
\begin{eqnarray}
\sigma_{n_Bp} = \sum_{m =0,\pm 1} \sigma_{n_Bp_m} = \sigma_{n_Bp_0}+2 \sigma_{n_Bp_1}\ .
\label{sigma_2CDR}
\end{eqnarray}
Here, $\sigma_{n_Bp_1}$ can be obtained from Eq.~\eqref{sigma_2CDR_0_final} by the replacements $\mathcal{B}_0({\bf R})\to \mathcal{B}_1({\bf R}) = \frac{9}{2}(\rho_x^2+\rho_y^2)\rho_z^2$ and $\Gamma_a^{(0)}\to\Gamma_a^{(1)}$. Note that interference terms between the three transition amplitudes $\mathcal{S}_{2p_m}$ do not contribute to the total, angle-integrated 2CDR cross section in Eq.~\eqref{sigma_2CDR}.

Besides, we have supposed that the electron is incident along a specific direction. If, instead, an average over the angles of incidence at fixed electron energy is taken, the averaged cross sections adopt the characteristic form
\begin{eqnarray}
\overline{\sigma}_{n_Bp_m} = \frac{\pi}{p^2}\,\frac{\Gamma_r^{(1s)}\Gamma_a^{(m)}}{\Delta^2+\frac{1}{4}\Gamma^2}\ .
\label{sigma_2CDR_0_av}
\end{eqnarray}
This can be seen by combining Eq.~\eqref{sigma_2CDR_0} with Eq.~\eqref{Gamma_Auger}. Similarly to Eq.~\eqref{sigma_2CDR}, the total 2CDR cross section, averaged over electron directions, is the sum of the individual contributions \eqref{sigma_2CDR_0_av} from the different intermediate states.

\section{Results and Discussion}

In the following we illustrate our findings on 2CRS and 2CDR by some examples. For definiteness, we shall assume throughout that the interatomic separation vector ${\bf R}=R{\bf e}_z$ lies along the momentum ${\bf p}=p\,{\bf e}_z$ of the incident electron which is captured by a proton ($Z_A=1$) to form a hydrogen atom at center $A$. Another light atomic system (H, He or He$^+$) represents center $B$.

\subsection{Properties of 2CRS}

\begin{figure}[b]  
\vspace{-0.25cm}
\begin{center}
\includegraphics[width=0.5\textwidth]{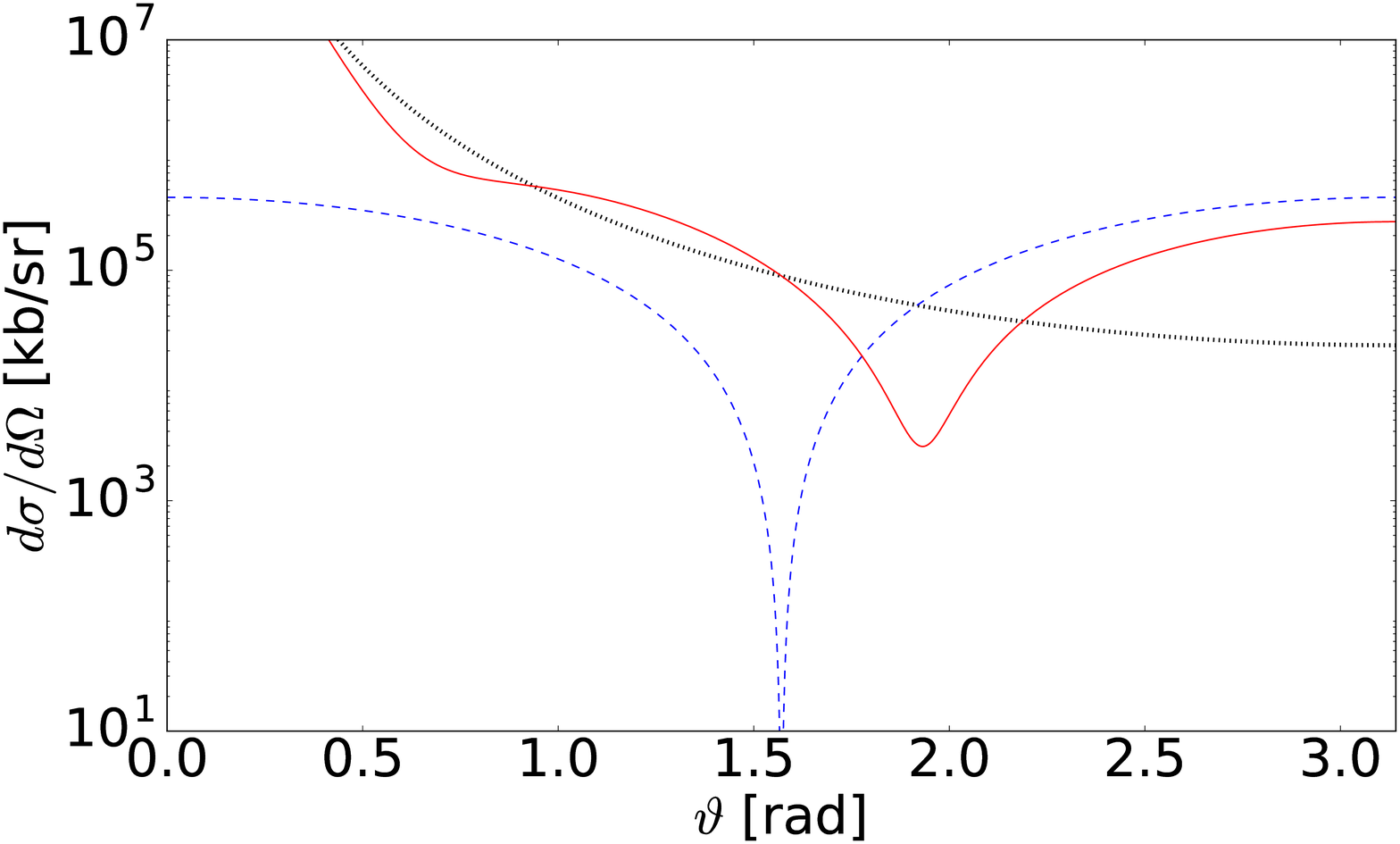}
\end{center}
\vspace{-0.5cm} 
\caption{Differential cross sections for Rutherford scattering from a proton and for 2CRS from a H$^+$-He system. In the latter case, the incident electron is captured to the $1s$ ground state of hydrogen, with simultaneous excitation of the helium atom to the $1s2p$ state. The atoms are separated by $R=5\,\mbox{\AA}$ along the $z$ axis. The incident electron energy is resonant with $\varepsilon_p = 7.6$\,eV. The dashed blue line shows the (incoherent) 2CRS cross section according to Eq.~\eqref{CSz}. For comparison, the dotted black line illustrates the cross section for Rutherford scattering from an isolated proton at the same electron energy [see Eq.~\eqref{CS_R}]. The solid red line shows the full cross section where the respective amplitudes have been added coherently [see Eq.~\eqref{CS_int}].}
\label{figure2}
\end{figure}

First we consider 2CRS with capture to the 1$s$ ground state of hydrogen, in the presence of a neutral helium atom at distance $R=5\,\mbox{\AA}$. During the process, the latter is excited to the $1s2p$ ($^1P^0$) state, which has an excitation energy of $\epsilon_1-\epsilon_0\approx 21.2$\,eV. For simplicity, we describe the helium atom in an approximate way as a hydrogenlike one-electron system with effective nuclear charge $Z_B=1.44$, which yields the correct excitation energy. The incident electron energy is chosen to fulfill the resonance condition. Figure~\ref{figure2} shows the corresponding 2CRS cross section from Eq.~\eqref{CSz} by the dashed blue line. It exhibits a characteristic dipole pattern and largely exceeds the cross section for Rutherford scattering (dotted black line) under backward angles. The relative enhancement factor at $\vartheta=\pi$ amounts to about 20, in agreement with the formula $(d\sigma_{2p_0}/d\Omega)/(d\sigma_{\rm R}/d\Omega) = 72\varepsilon_p$ found in Sec.~\ref{1s}. Note that at the chosen internuclear distance, the relation $\Gamma_a \gg \Gamma_r$ holds.

The full process, including quantum interference between both scattering pathways, is displayed by the solid red line in Fig.~\ref{figure2}. It exhibits a pronounced minimum close to $\theta\approx 1.9$\,rad where the separate contributions from 2CRS and Rutherford scattering cross. Under backward angles, the enhancement is still quite strong, but somewhat reduced, indicating destructive interference between both channels. The latter can be understood by noting that, for backward scattering ($\vartheta=\pi$), the interference phase becomes $\Phi(\pi) = 2\arctan(\nu)$, so that $\sin\Phi(\pi) = 2\nu/(1+\nu^2)>0$. In the considered geometry, the functions in Eq.~\eqref{B01} reduce to $B_0 = 4\cos\vartheta = -4$, $B_1=0$, so that $C(\vartheta,\phi)$ in Eq.~\eqref{C} becomes negative since $A(\vartheta)$ is always positive. Therefore, the interference in Eq.~\eqref{CS_int} is destructive for $\vartheta=\pi$.

\begin{figure}[h]  
\vspace{-0.25cm}
\begin{center}
\includegraphics[width=0.5\textwidth]{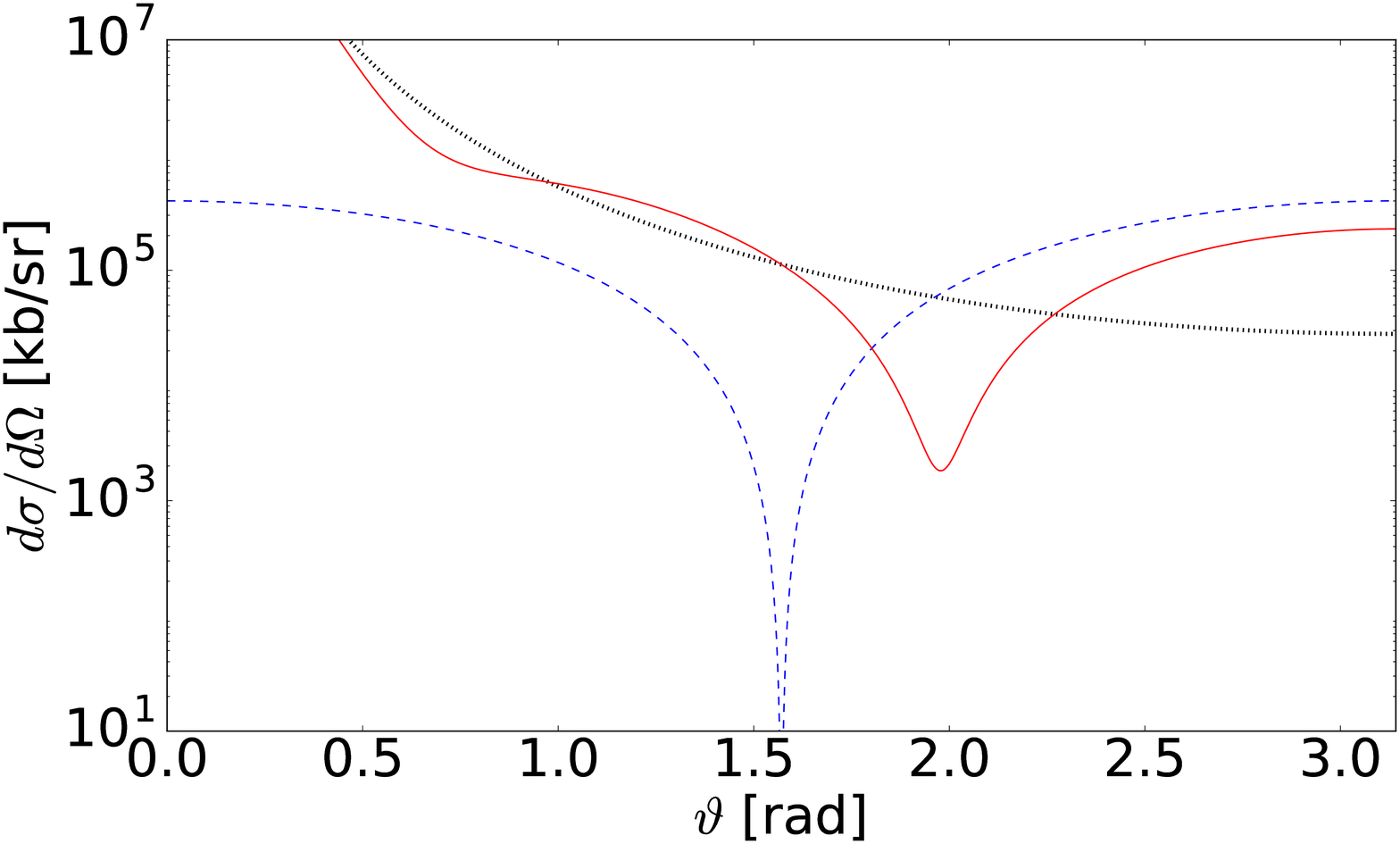}
\includegraphics[width=0.5\textwidth]{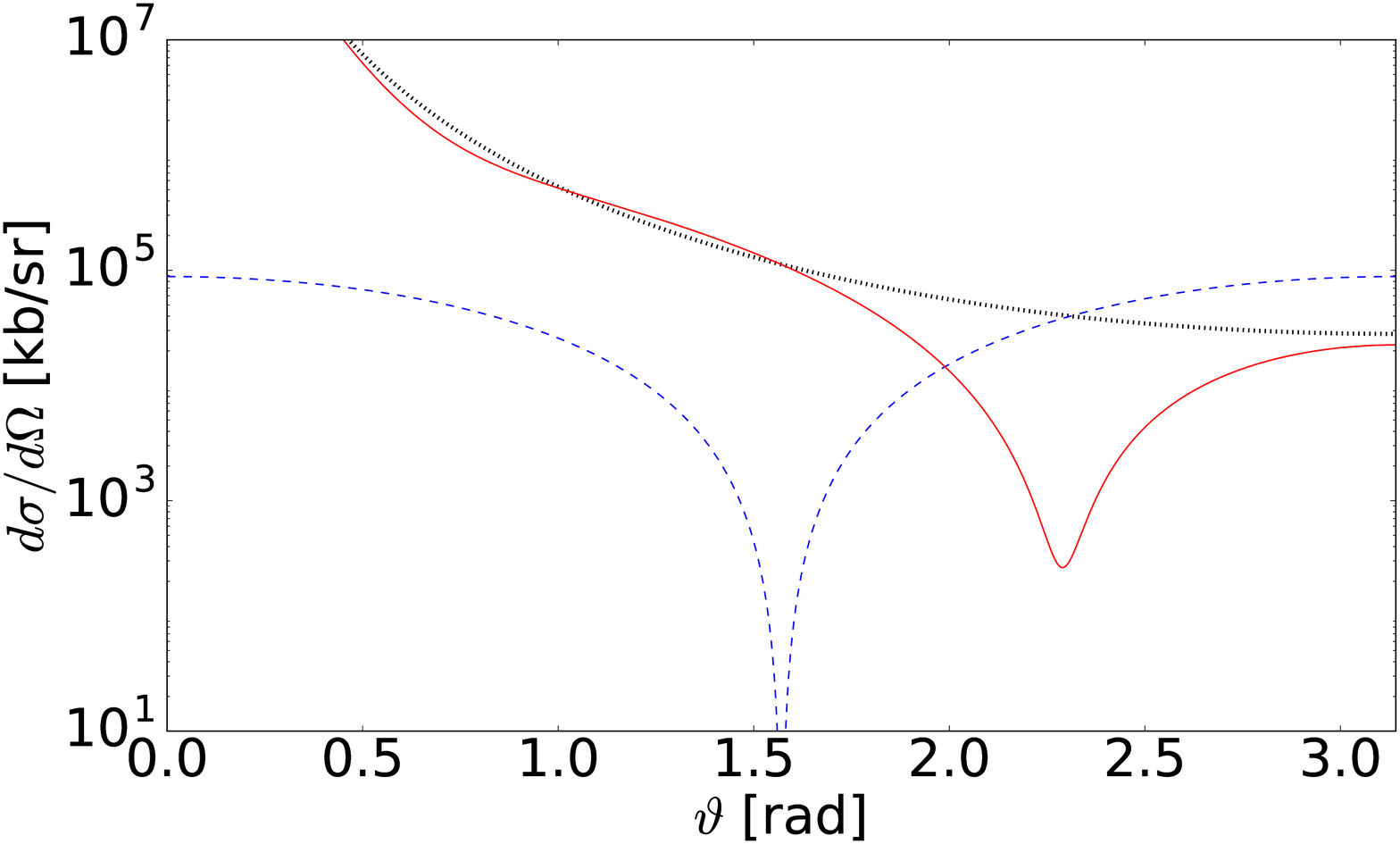}
\end{center}
\vspace{-0.5cm} 
\caption{Same as Fig.~\ref{figure2}, but in the 2CRS process the electron is captured to the $2s$ state in the hydrogen atom forming center $A$, while another hydrogen atom -- representing center $B$ -- is simultaneously excited to the $2p_0$ state. The energy of the incident electron is in exact resonance with $\varepsilon_p = 6.8$\,eV. Top: $R=10\,\mbox{\AA}$, bottom: $R=15\,\mbox{\AA}$.}
\label{figure3}
\end{figure}

As we saw in Sec.~III.A, the 2CRS cross section is very sensitive to the energy of the involved transitions. The latter can be strongly reduced when the helium atom at center $B$ is replaced by a hydrogen atom, which is excited from the $1s$ to the $2p_0$ state, with $\epsilon_1-\epsilon_0\approx 10.2$\,eV. The resonance condition can be met by an electron which is incident with momentum $p=1/\sqrt{2}$\,a.u. and captured by the proton at center $A$ to the $2s$ state of hydrogen. The corresponding 2CRS cross section is shown in Fig.~\ref{figure3}. In the top panel we see that a similar enhancement under backward angles like in Fig.~\ref{figure2} can be achieved, but at a twice as large interatomic separation. A relative enhancement factor of $\approx 14$ is reached for $\vartheta=\pi$. It is slightly reduced as compared with the plateau value $72\varepsilon_p\approx 18$ by the squared ratio of Auger width to total width, which amounts to $\Gamma_a^2/\Gamma^2\approx 0.8$ at $R=10\,\mbox{\AA}$.

The bottom panel in Fig.~\ref{figure3} illustrates that, at an increased internuclear distance of $15\,\mbox{\AA}$,  the full process including interference does not show enhancement under backward angles as compared with Rutherford scattering anymore. However, a very pronounced minimum around $\vartheta\approx 2.3$\,rad appears due to the influence of 2CRS.

It is interesting to compare our findings with results on the single-center process of elastic electron  scattering from ions which carry at least one bound electron. In this situation, there are also bound-state resonances which can have pronounced influence on the scattering cross section. A combined experimental and theoretical study of elastic scattering of electrons at 16 eV energy from Ar$^+$ ions was performed in \cite{Gribakin}. Qualitatively, our findings on 2CRS closely resemble the results shown, for example, in Fig.~3 therein. In both cases, the full cross section including interference effects has a similar shape.  In particular, it passes through a minimum at intermediate angles due to destructive interference and shows pronounced enhancement under backward angles as compared with Rutherford scattering.

\subsection{Scaling behavior of 2CDR}

\begin{figure}[b]  
\vspace{-0.25cm}
\begin{center}
\includegraphics[width=0.5\textwidth]{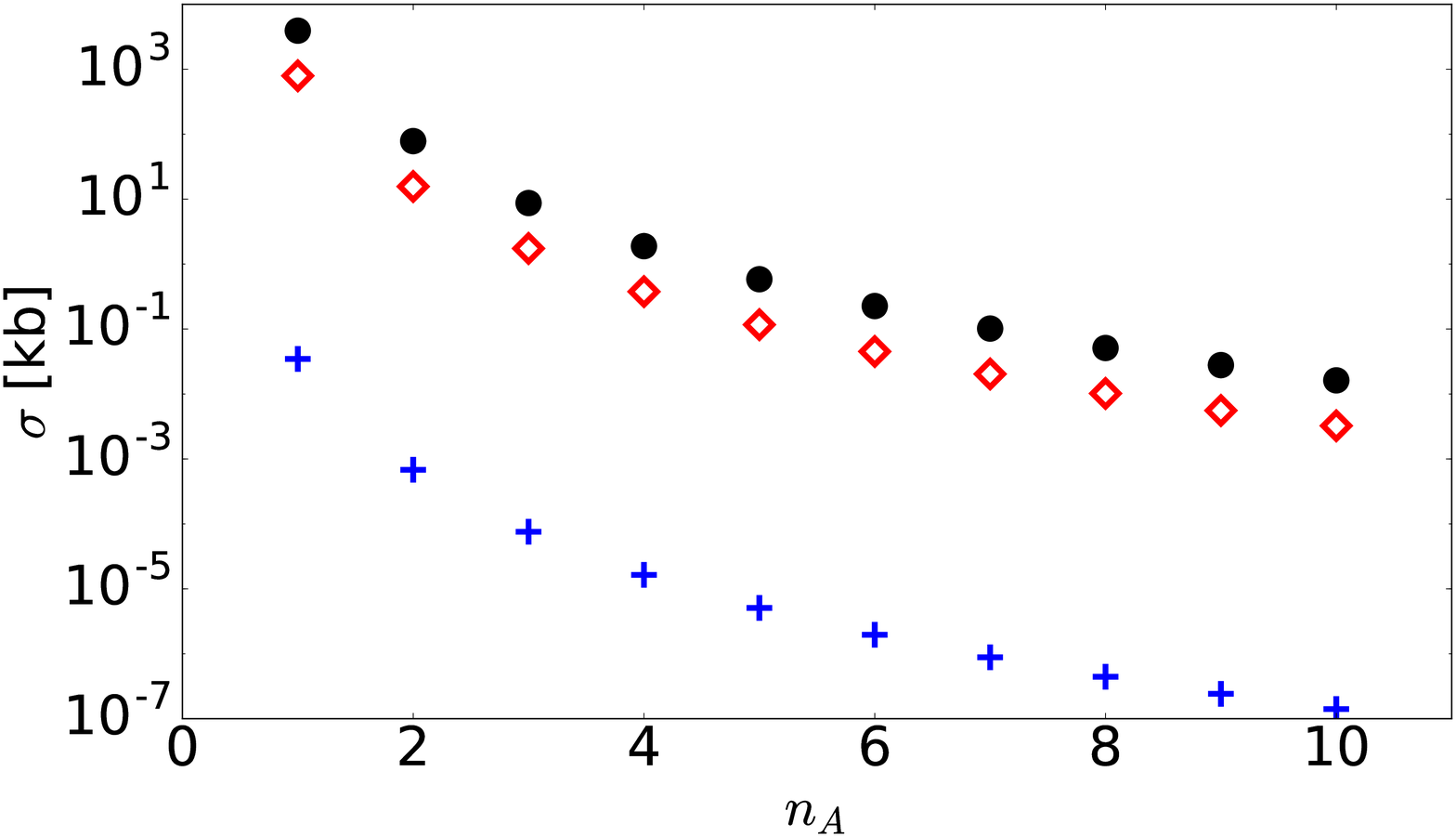}
\end{center}
\vspace{-0.5cm} 
\caption{Dependence of the 2CDR cross section on the principal quantum number $n_A$. An incident electron recombines with a proton to form a hydrogen atom in the presence of a neighboring He$^+$ ion. The electron is captured to the $n_As$ state of hydrogen, with simultaneous excitation of the He$^+$ ion to the $2p_0$ state. The atoms are separated by $R=10\,\mbox{\AA}$ along the $z$ axis. The black circles refer to the case of exact resonance, where for each value of $n_A$ the incident energy is chosen to give $\Delta=0$ [see Eq.~\eqref{sigma_2CDR_0_final}]. Instead, the red diamonds show the corresponding results for a non-zero detuning of $\Delta=6.6\times10^{-6}$\,eV. For comparison, the crosses show the cross section for radiative recombination of the electron with an isolated proton [see Eq.~\eqref{CS_RR}].}
\label{figure4}
\end{figure}

Now we turn to 2CDR. First we consider an electron which is captured to the $n_As$ state in hydrogen, with simultaneous excitation of a neighboring He$^+$ ion from the ground to the $2p_0$ state. Figure~\ref{figure4} shows the $n_A$-dependence of the corresponding 2CDR cross section. Note that, for every value of $n_A$, the energy of the incident electron has been adjusted to either fulfill the resonance condition exactly ($\Delta=0$, black circles) or to yield a fixed detuning ($\Delta\approx\Gamma_r$, red diamonds). For comparison, the cross section of single-center radiative recombination is shown (blue crosses). All curves fall off monotoneously, showing a dependence like $n_A^{-3}$ to a good approximation. This kind of scaling law, which is well-known for radiative recombination, is thus shared by the 2CDR cross section. Besides, Fig.~\ref{figure4} shows that, on the resonance or very close to it, 2CDR can strongly dominate over radiative recombination. For the present parameters, the relative enhancement amounts to five orders of magnitude, in agreement with Eq.~\eqref{2CDR_ratio_2p}.

\begin{figure}[b]  
\vspace{-0.25cm}
\begin{center}
\includegraphics[width=0.5\textwidth]{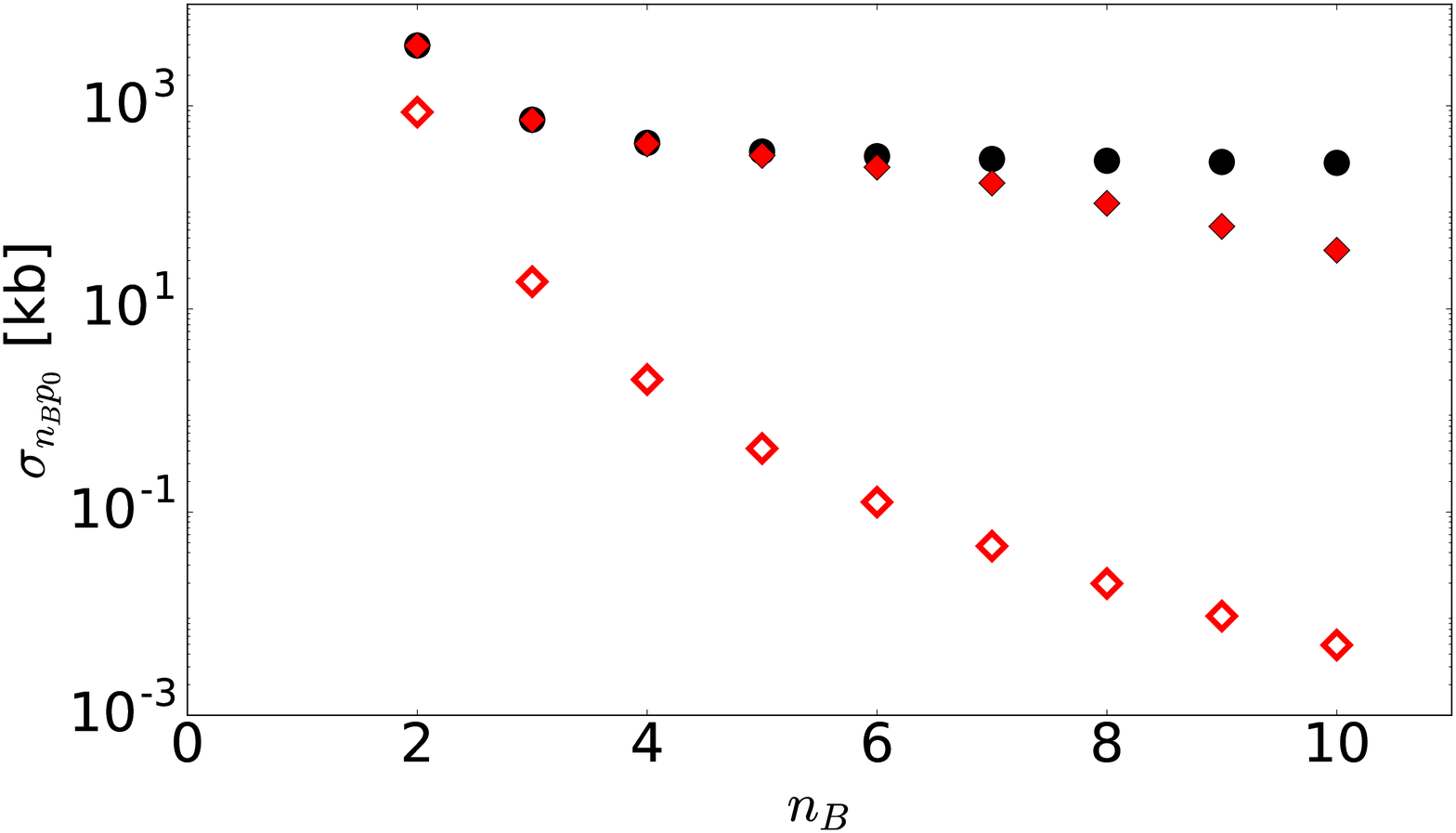}
\end{center}
\vspace{-0.5cm} 
\caption{Dependence of the 2CDR cross section on the principal quantum number $n_B$. As in Fig.~\ref{figure4}, an incident electron recombines with a proton to form a hydrogen atom in the presence of a neighboring He$^+$ ion. The electron is captured to the $1s$ ground state of hydrogen, with simultaneous excitation of the He$^+$ ion to the $n_Bp_0$ state. The atoms are separated by $R=10\,\mbox{\AA}$ along the $z$ axis. While the black circles refer to the case of exact resonance, where for each value of $n_B$ the incident energy is chosen to give $\Delta=0$, the red filled (open) diamonds show the corresponding results for non-zero detuning of $\Delta=6.6\times 10^{-8}$\,eV ($\Delta=6.6\times 10^{-6}$\,eV) [see Eq.~\eqref{sigma_2CDR_0_final}].}
\label{figure5}
\end{figure}

Figure~\ref{figure5} displays the dependence of 2CDR on the quantum number $n_B$, for vanishing and non-zero detunings. Here we assume that the incident electron is captured to the $1s$ ground state of hydrogen at center $A$, while the He$^+$ ion is excited to the $n_Bp_0$ state. For each value of $n_B$, the incident electron energy is adjusted, accordingly. In the case of exact resonance ($\Delta=0$, black circles) we obtain a result which may appear surprising at first sight. The 2CDR cross section first decreases, but then saturates rather quickly for larger values of $n_B$. This striking feature can be understood by inspecting Eq.~\eqref{sigma_2CDR_0} which implies that $\sigma_{n_Bp_0}\propto \Gamma_r^{(1s)}|\mathcal{M}({\bf p})|^2/(p\,\Gamma^2)$. From Eqs.~\eqref{M_gen}, \eqref{Gamma_rad} and \eqref{Gamma_Auger_0} we know the leading-order dependencies $|\mathcal{M}({\bf p})|^2$, $\Gamma_r^{(1s)}$, $\Gamma_a\sim n_B^{-3}$ which, accordingly, drop out from the cross section. What remains is the momentum, whose value for exact resonance follows from $p_{\rm res}^2 = 2\left[\varepsilon_0 + \epsilon_0\left( \frac{1}{n_B^2}-1\right)\right]$ and, thus, becomes approximately constant for $n_B\gg 1$. Combining these arguments, we see that the 2CDR cross section indeed saturates for large values of $n_B$, provided the resonance condition is exactly met.

In case of non-zero detuning, the 2CDR cross section shows a very different behavior. If the detuning is rather large ($\Delta>\Gamma$ for $n_B > 2$), the cross section rapidly falls off and does not show saturation, as depicted by red open diamonds in Fig.~\ref{figure5}. The scaling law to leading order then reads $\sigma_{n_Bp_0}\propto \Gamma_r^{(1s)}|\mathcal{M}({\bf p})|^2\sim n_B^{-6}$, in agreement with our numerical results. In the case of small detuning (red filled diamonds in Fig.~\ref{figure5}), the 2CDR cross section shows an intermediate behavior. It first decreases, passes a plateau region and eventually falls off when $n_B$ grows. This is caused by the fact that the total width $\Gamma$ decreases with $n_B$ and, thus, at some point falls below the detuning $\Delta$ which is assumed to be fixed. Then, the 2CDR cross section approaches again a scaling with $n_B^{-6}$.

\begin{figure}[b]  
\vspace{-0.25cm}
\begin{center}
\includegraphics[width=0.5\textwidth]{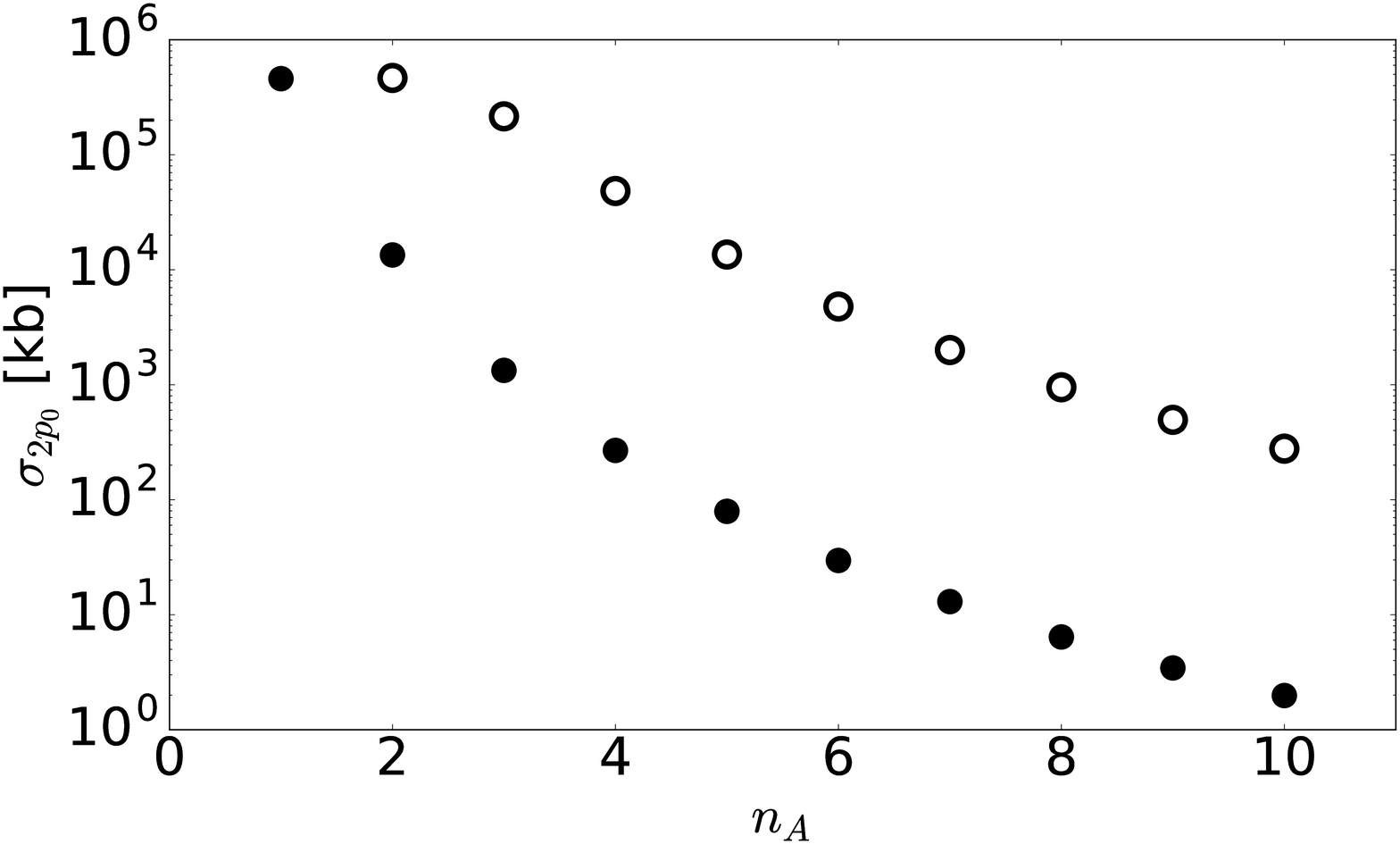}
\end{center}
\vspace{-0.5cm} 
\caption{Dependence of 2CDR on the principal quantum number $n_A$ for two different atomic systems. In both cases, an incident electron recombines with a proton at center $A$, forming a hydrogen atom in the $n_As$ state. The filled circles refer to a neutral helium atom as center $B$ which is excited to the $1s2p$ state, whereas the open circles show our results for a neighboring hydrogen atom which is excited to the $2p_0$ state. The atoms are separated by $R=10\,\mbox{\AA}$ along the $z$ axis and the incident electron energies are always chosen to be in exact resonance.}
\label{figure6}
\end{figure}

It is interesting to note that for some specific electron energies, there is more than one combination of $n_As$ and $n_Bp$ states which allows to fulfill the resonance condition. For such energies, the electron can be either captured to a deeply bound state in atom $A$ with excitation of a high-lying state in atom $B$, or the capture proceeds to a higher lying state in atom $A$ accompanied by a transition with correspondingly lower excitation energy in atom $B$. Let us take the two-center system with $Z_A=1$,  $Z_B=2$ considered in this section as an example. For each integer $n\in\mathbb{N}$, the combination $(n_A,n_B)=(1,2n+2)$ is associated with the same resonant incident electron energy as the combination $(n_A',n_B')=(n+1,2)$, because the sum $\varepsilon_0+\epsilon_1$ has the same value. Assuming exact resonance ($\Delta=0$), we find that the first channel gives a slightly smaller 2CDR cross section for $n=1$, but largely dominates over the second channel when $n$ increases \cite{Jacob}. The reason is that the cross section falls off steeply with $n_A^{-3}$, whereas it is much more weakly dependent on $n_B$ (see Figs.~\ref{figure4} and \ref{figure5}).

Finally, Fig.~\ref{figure6} shows 2CDR cross sections when, instead of a He$^+$ ion, a neutral helium or a hydrogen atom forms center $B$. As in Sec.~IV.A the helium atom is described as an effective one-electron system with $Z_B=1.44$. Since the excitation energy in He is about half the value in He$^+$, the corresponding 2CDR cross sections in Fig.~\ref{figure6} (filled circles) lie two orders of magnitude higher than in Fig.~\ref{figure4}. The 2CDR cross section becomes another two orders of magnitude larger in the presence of a neighboring H atom (open circles), whose excitation energy is about half the value in neutral He. Note that the corresponding cross section for $n_A=2$ is damped by the fact that here $\Gamma_a\gg\Gamma_r$ holds. Instead, for the large values of $n_A$ we have $\Gamma_a\ll\Gamma_r$, giving rise to the $n_A^{-3}$ dependence which was already observed in Fig.~\ref{figure4}.

We point out that Figures~\ref{figure4}--\ref{figure6} are meant to illustrate the general dependence of 2CDR on the principal quantum numbers in the participating atoms. Our results rely on the assumption that the electron wave functions at both centers do not overlap with each other. While, at the chosen interatomic distance of $R=10\,\mbox{\AA}$, this condition is satisfied to a good approximation for the small values of $n_{A,B}$, it is not anymore for the larger ones. Nevertheless, by exploiting the proportionality $\sigma\sim R^{-6}$, the cross sections shown in the figures can easily be extrapolated to larger $R$ values, where the assumption of non-overlapping wave functions is fulfilled.

\subsection{Comparison between 2CRS and 2CDR}
From Eqs.~\eqref{CS_tot} and \eqref{sigma_2CDR_0} we can read off that the ratio between the cross sections for 2CRS and 2CDR amounts to $\Gamma_a/\Gamma_r^{(1s)}$. Moreover, exactly on the resonance, we have the proportionalities $\sigma_{\chi_1}\sim \lambda_p^2\, \Gamma_a^2/(\Gamma_r+\Gamma_a)^2$ for the 2CRS cross section and $\sigma_{n_Bp_m}\sim \lambda_p^2\, \Gamma_r^{(1s)}\Gamma_a/(\Gamma_r+\Gamma_a)^2$ for the 2CDR cross section. Here, $\lambda_p$ denotes the de-Broglie wavelength of the incident electron. Accordingly, in the range of small interatomic distances where $\Gamma_r < \Gamma_a$ holds, we see that the 2CRS process becomes $R$-independent, with $\sigma_{\chi_1}\sim \lambda_p^2$. In contrast, for large values of $R$ with $\Gamma_a \ll \Gamma_r$, we obtain $\sigma_{\chi_1}\sim \lambda_p^2\, (\Gamma_a/\Gamma_r)^2$ which is smaller than the 2CDR cross section $\sigma_{n_Bp_m}\sim \lambda_p^2\, \Gamma_a/\Gamma_r$.

When the radiative width $\Gamma_r$ and two-center Auger width $\Gamma_a$ have similar values, the cross sections for 2CRS and 2CDR are of the same order of magnitude. Nevertheless, the ratio to the corresponding single-center process was found to be very large for 2CDR, whereas only a rather moderate enhancement for certain angles was found for 2CRS. The reason why the relative enhancement is much larger for 2CDR is due to the fact that, in this case, the comparison is made with the relatively weak process of radiative recombination. In contrast, 2CRS must be compared with Rutherford scattering, which represents a very strong interaction process between charged particles without coupling to the radiation field.

\subsection{Comparison with single-center processes}
Resonant electron scattering (RS) and dielectronic recombination (DR) are well-known to occur as single-center processes in isolated ions, where they are driven by inner-ionic electron correlations \cite{AMueller}. One can compare, in very general terms, the strengths of these processes with 2CRS and 2CDR. 

In analogy to Eq.~\eqref{sigma_2CDR_0}, the cross section of single-center DR is proportional to $\sigma_{\rm DR}\sim \lambda_p^2\,\tilde{\Gamma}_a\Gamma_r/[\Delta^2 + (\Gamma_r+\tilde{\Gamma}_a)^2]$. Here, $\tilde{\Gamma}_a$ is used to denote the single-center Auger width, in order to distinguish it from the two-center Auger width $\Gamma_a$, which is in general much smaller. For the radiative width we do not introduce a new symbol because it represents a single-center quantity and is, thus, of similar size in both cases. For light atomic systems, where $\tilde{\Gamma}_a\gg\Gamma_r$ holds, we obtain $\sigma_{\rm DR}\sim \lambda_p^2\Gamma_r/\tilde{\Gamma}_a$ on the resonance. The ratio with 2CDR thus becomes $\sigma_{n_Bp_m}/\sigma_{\rm DR}\sim \Gamma_a\tilde{\Gamma}_a/\Gamma_r^2$, assuming that $\Gamma_a \ll \Gamma_r$ (see Sec.~IV.C). From this relation we can infer that, in terms of the resonant value of the cross section, 2CDR can indeed compete with and even exceed single-center DR.

Similarly, we have $\sigma_{\rm RS}\sim \lambda_p^2\,\tilde{\Gamma}_a^2/[\Delta^2 + (\Gamma_r+\tilde{\Gamma}_a)^2]$ for single-center RS, in analogy with Eq.~\eqref{CS_tot}. Due to $\tilde{\Gamma}_a\gg\Gamma_r$, here we obtain $\sigma_{\rm RS}\sim \lambda_p^2$ on the resonance. The same result we found for 2CRS in the range of interatomic distances where $\Gamma_a \gg \Gamma_r$ holds (see Sec.~IV.C), so that the ratio becomes $\sigma_{\chi_1}/\sigma_{\rm RS}\sim 1$. Instead, in the opposite case $\Gamma_a \ll \Gamma_r$, the ratio attains a small value of $\sigma_{\chi_1}/\sigma_{\rm RS}\sim (\Gamma_a/\Gamma_r)^2\ll 1$.

So far, our discussion has refered to the case of exact resonance. One has to keep in mind, though, that in an experiment the incident electron beam will have a certain energetic width $\delta\varepsilon$, corresponding to a distribution of energies around the resonant value. Since typically $\delta\varepsilon\gg \Gamma$, only a small fraction $\sim\Gamma / \delta\varepsilon\ll 1$ of all electrons can effectively contribute to 2CRS and 2CDR. When Eqs.~\eqref{sigma_2CDR_0} and \eqref{CS_tot} are integrated over incident energies, we obtain the quantities $\langle\sigma_{n_Bp_0}\rangle \sim \lambda_p^2\,\Gamma_a\Gamma_r/ (\Gamma_r+\Gamma_a)$ and $\langle\sigma_{\chi_1}\rangle \sim \lambda_p^2\,\Gamma_a^2/(\Gamma_r+\Gamma_a)$, which describe the areas under the resonance peaks. For 2CDR at $\Gamma_a < \Gamma_r$, this yields $\langle\sigma_{n_Bp_0}\rangle \sim \lambda_p^2\,\Gamma_a$. The corresponding quantity for single-center DR becomes $\langle\sigma_{\rm DR}\rangle \sim \lambda_p^2\,\Gamma_r$ which is larger, but not necessarily much larger than $\langle\sigma_{n_Bp_0}\rangle$. 

In contrast, for 2CRS we obtain $\langle\sigma_{\chi_1}\rangle \sim \lambda_p^2\,\Gamma_a$ (when $\Gamma_a>\Gamma_r$) or even $\langle\sigma_{\chi_1}\rangle \sim \lambda_p^2\,\Gamma_a^2/\Gamma_r$ (when $\Gamma_a<\Gamma_r$), which both are substantially smaller than $\langle\sigma_{\rm RS}\rangle \sim \lambda_p^2\,\tilde{\Gamma}_a$ for single-center RS. In this context, it is important to note that the two-center Auger width $\Gamma_a$ of Eq.~\eqref{Gamma_Auger} is, in general, several orders of magnitude smaller than single-center Auger widths $\tilde{\Gamma}_a$ in isolated ions. For example, in our Figs.~\ref{figure2} and \ref{figure3} we have $\Gamma_a\sim \mu$eV, whereas in Ref.~\cite{Gribakin} a typical value of $\tilde{\Gamma}_a=50$\,meV was applied. Therefore, while 2CRS can reach similar peak values of the resonant cross section as single-center RS, it cannot compete in terms of the energy-integrated resonance strength.

There is a physically intuitive reason why the energy-integrated resonance strengths of 2CDR and single-center DR can be comparable, whereas for 2CRS and single-center RS this is not the case. At first sight, one would expect single-center DR to be much more probable, because the interaction between two electrons is much stronger, if they are spatially confined to the volume of an ion, than if they belong to two different atomic systems which lie rather far apart. However, a strong electron-electron interaction also leads to a high probability for Auger decay of the autoionizing state formed after electron capture. Therefore, the majority of captured electrons is reemitted and does not contribute to DR in isolated (light) atoms or ions where $\tilde{\Gamma}_a\gg\Gamma_r$ holds. For 2CDR the situation is different because, at not too small interatomic distances, the two-center Auger width is very small, $\Gamma_a\ll\Gamma_r$. Therefore, practically all electrons, which are captured, also contribute to 2CDR because the autoionizing state stabilizes radiatively in most cases. Therefore, 2CDR can be competitive with single-center DR despite the much weaker electron-electron interaction involved. 

For 2CRS the above line of argument does not work. Here, a relatively large radiative width is not advantageous, because relaxation of the autoionizing state through spontaneous radiative decay does not contribute to electron scattering. Instead, 2CRS relies on the narrow two-center Auger width, which results from the rather weak interatomic electron-electron interaction and which is much smaller than single-center Auger widths in the relevant range of interatomic distances. Therefore, 2CRS falls behind single-center RS when their energy-integrated resonance strengths are compared.

\section{Conclusion}
Electron-impact induced processes in two-center atomic systems, which rely on resonant electron-electron interactions, have been considered. 
The process of 2CRS, where an electron scatters resonantly from a two-center system, was introduced and shown to considerably modify the scattering dynamics at interatomic distances up to $\sim 1$\,nm, as compared with the well-known Rutherford scattering from a single charge center. In particular, for light atomic systems, 2CRS may enhance the scattering under backward angles by an order of magnitude. Due to quantum interference, it can also significantly affect scattering at intermediate angles. For small angles, Rutherford scattering dominates due to its divergent behavior in forward direction. 

In comparison with resonant scattering from single-center atomic systems, 2CRS can compete in terms of the resonant value of the cross section. But it falls behind when an integral over incident energies is taken, because its resonance width is much more narrow. In a forthcoming study we plan to include also the elastic electron scattering from the neighboring atom into the description of 2CRS.

Furthermore, dielectronic recombination with a two-center system was studied. By calculating the corresponding cross section for capture into general $n_As$ states of an ion with simultaneous excitation to a $n_Bp$ state of a neighboring atom, the scaling behavior of 2CDR with the principal quantum numbers was obtained. A general proportionality with $n_A^{-3}$ was found, like it is known for single-center radiative recombination. Exactly on resonance, the 2CDR cross section quickly saturates to a constant value when $n_B$ grows. In contrast, a more complex behavior was found in the detuned case, where the cross section asymptotically scales with $n_B^{-6}$. 

For incident electron energies close to the resonance and interatomic distances up to few nanometers, the 2CDR cross section can exceed the cross section for single-center radiative recombination by several orders of magnitude. It can also compete with typical cross sections of single-center DR, in terms of both the peak value on the resonance and, remarkably, also the energy-integrated resonance strength. The latter result may be understood intuitively by noting that, contrary to single-center DR, almost all electron capture events contribute to 2CDR. This is because the populated autoionizing state relaxes preferably via spontaneous radiative decay, since the radiative width greatly exceeds the two-center Auger width at sufficiently large interatomic distances.

\section*{Acknowledgement}
This study has been performed within the projects MU 3149/4-1 and VO 1278/4-1 funded by the German Research Foundation (DFG). The first two authors of the present paper (A.~E. and A.~J.) contributed equally.
We thank G.~Gribakin for useful conversations on 2CDR and
L.~Silletti for her help at the onset of this study.


\end{document}